\documentclass[review]{elsarticle}

\usepackage{lineno,hyperref}
\usepackage{amsmath}
\usepackage{color}
\modulolinenumbers[5]

\journal{Physica A}

\begin{document}

\begin{frontmatter}

\title{Boundary Effects on Population Dynamics in Stochastic Lattice 
       Lotka--Volterra Models}

\author{Bassel Heiba, Sheng Chen, and Uwe C. T\"auber}
\address{Department of Physics (MC 0435) and Center for Soft Matter and Biological 
	Physics, Robeson Hall, 850 West Campus Drive, Virginia Tech, Blacksburg, 
	Virginia 24061, USA}
\ead{bassel5@vt.edu, csheng@vt.edu, tauber@vt.edu}


\begin{abstract}
We investigate spatially inhomogeneous versions of the stochastic Lotka--Volter\-ra 
model for predator-prey competition and coexistence by means of Monte Carlo 
simulations on a two-dimensional lattice with periodic boundary conditions. To 
study boundary effects for this paradigmatic population dynamics system, we employ
a simulation domain split into two patches: Upon setting the predation rates at two 
distinct values, one half of the system resides in an absorbing state where only 
the prey survives, while the other half attains a stable coexistence state wherein
both species remain active. At the domain boundary, we observe a marked enhancement
of the predator population density. The predator correlation length displays a 
minimum at the boundary, before reaching its asymptotic constant value deep in the 
active region. The frequency of the population oscillations appears only very 
weakly affected by the existence of two distinct domains, in contrast to their 
attenuation rate, which assumes its largest value there. We also observe that 
boundary effects become less prominent as the system is successively divided into 
subdomains in a checkerboard pattern, with two different reaction rates assigned to
neighboring patches. When the domain size becomes reduced to the scale of the 
correlation length, the mean population densities attain values that are very 
similar to those in a disordered system with randomly assigned reaction rates drawn
from a bimodal distribution. 
\end{abstract}

\begin{keyword}
stochastic population dynamics \sep Lotka--Volterra predator-prey competition 
\sep spatial inhomogeneities \sep boundary effects \sep quenched disorder
\end{keyword}

\end{frontmatter}


\section{Introduction}

Due to its wide range of applications and relative simplicity, variants of the 
Lotka--Volterra predator-prey competition model represent paradigmatic systems to 
study the emergence of biodiversity in ecology, noise-induced pattern formation in 
population dynamics and (bio-)chemical reactions, and phase transitions in 
far-from-equilibrium systems. In the classical deterministic Lotka--Volterra 
model~\cite{a1920, v1926}, two coupled mean-field rate equations describe the 
population dynamics of a two-species predator-prey system, whose solutions display 
periodic non-linear oscillations fully determined by the system's initial state. 
Yet the original mean-field Lotka--Volterra rate equations do not incorporate 
demographic fluctuations and internal noise induced by the stochastic reproduction 
and predation reactions in coupled ecosystems encountered in nature. In a series of 
analytical~\cite{hnak1992}--\cite{r1999} and numerical simulation 
studies~\cite{agf1999}--\cite{ct2016}, the population dynamics of several stochastic spatially extended lattice 
Lotka--Volterra model variants was found to substantially differ from the mean-field
rate equation predictions due to stochasticity and the emergence of strong 
spatio-temporal correlations: Both predator and prey populations oscillate 
erratically, and do not return to their initial densities; the oscillations are 
moreover damped and asymptotically reach a quasi-stationary state with both 
population densities finite and constant on one- or two-dimensional square 
lattices~\cite{agf1999}, whereas damping appears absent or is very weak in three 
dimensions~\cite{a1999}. Very similar dynamical properties are observed in other 
two-dimensional model variants, including a predator-prey system with added prey 
food supply and cover~\cite{ma2001}, and implementations on a triangular 
lattice~\cite{maa2002}. In a non-spatial setting, the persistent non-linear 
oscillations can be understood through resonantly amplified demographic 
fluctuations~\cite{mn2005}. Local carrying capacity restrictions, representing 
limited resources in nature, can be implemented in lattice simulations by 
constraining the number of particles on each 
site~\cite{ae1999, ad2000, rae2000, miu200602, mit2006, mmu2007}. These local 
occupation number restrictions cause the emergence of a predator extinction 
threshold and an absorbing phase, wherein the predator species ultimately disappears
while the prey proliferate through the entire system. Upon tuning the reaction 
rates, one thus encounters a continuous active-to-absorbing state non-equilibrium 
phase transition whose universal features turn out to be governed by the directed 
percolation universality 
class~\cite{jt1994, nom1994, ad2000, rae2000, ma2001, miu200602, mit2006, ct2016}.

Biologically more relevant models should include spatial rate variability to 
account for environmental disorder. The population dynamics in a patch surrounded 
by a hostile foe~\cite{s1951, dns1999, md2008} is well represented by Fisher's 
model~\cite{f1937}, which includes diffusive spreading as well as a reaction term 
capturing interactions between individuals and with the environment. For the
stochastic Lotka--Volterra model, the influence of environmental rate variability 
on the population densities, transient oscillations, spatial correlations, and 
invasion fronts was investigated by assigning random reaction rates to different 
lattice sites~\cite{uu2008, uu2013}. Spatial variability in the predation rate 
results in more localized activity patches, a remarkable increase in the asymptotic
population densities, and accelerated front propagation. These studies assumed full
environmental disorder, as there was no correlation at all between the reaction 
rates on neighboring sites. 

In a more realistic setting, the system should consist of several domains with the 
environment fairly uniform within each patch, but differing markedly between the 
domains, e.g., representing different topographies or vegetation states. In our 
simulations, we split the system into several patches and assign different reaction
rates to neighboring regions. By tuning the rate parameters, we can force some 
domains to be in an absorbing state, where the predators go extinct, or 
alternatively in an active state for which both species coexist at non-zero 
densities. One would expect the influence of the boundary between the active and 
absorbing regions to only extend over a distance on the scale of the characteristic
correlation length in the system. In this work, we study the local population 
densities, correlation length, as well as the local oscillation frequency and 
attenuation, as functions of the distance from the domain boundary. As we 
successively divide the system further in a checkerboard pattern so that each patch
decreases in size, the population dynamics features quantitatively tend towards 
those of a randomly disordered model with reaction rates assigned to the lattice 
sites from a bimodal distribution.

\section{Model Description and Background}

The deterministic classical Lotka--Volterra model~\cite{a1920, v1926} is a set of
two coupled non-linear dynamical rate equations that on a mean-field approximation
level capture the following kinetic reactions of two species, respectively 
identified with predators $A$ and prey $B$:
\begin{equation}
\label{lvreac}
  A \overset{\mu}{\rightarrow} \emptyset \, , \quad
  A + B \overset{\lambda}{\rightarrow} A + A \, , \quad
  B \overset{\sigma}{\rightarrow} B + B \, .
\end{equation}
In these stochastic processes, $\mu$ corresponds to the spontaneous predator death 
rate, while $\sigma$ denotes the prey reproduction rate. Finally, $\lambda$ is the 
predation rate which describes the non-linear reaction through which the predator 
and prey species interact with each other. The simplified Lotka--Volterra model 
thus assumes that the prey population grows exponentially in the absence of 
predators, but becomes diminished with growing predator population. In the presence
of the prey, the predator population will increase with the prey population, but is 
subject to exponential decay once all prey are gone. The configuration with 
vanishing predator number represents an absorbing state for this system, since there
exists no stochastic reaction process that would allow recovery from it. For 
completeness, we mention that the total population extinction state of course 
represents another absorbing state. We also remark that one could add independent 
predator reproduction $A \rightarrow A + A$ (with rate $\sigma'$) and prey death 
$B \rightarrow \emptyset$ (rate $\mu'$) processes to the standard Lotka--Volterra 
kinetics (\ref{lvreac}). Yet this would induce no qualitative changes as long as 
$\sigma' < \mu$ and $\sigma > \mu'$; one then simply needs to replace $\mu$ with the
rate difference $\mu - \sigma'$, and $\sigma$ with $\sigma - \mu'$.

The associated rate equations, subject to mean-field mass action factorization for 
the non-linear predation process, and valid under well-mixed conditions for 
spatially homogeneous time-dependent particle densities $a(t)$ and $b(t)$, read
\begin{equation}
\label{lvmfeq}
  \dot a(t) = \lambda' a(t) b(t) - \mu' a(t) \, , \quad 
  \dot b(t) = - \lambda' a(t) b(t) + \sigma' b(t) \, ,
\end{equation}
with continuum reaction rates $\lambda'$, $\mu'$, and $\sigma'$. Since there exists 
a conserved first integral (Lyapunov function) 
$K = \lambda' (a + b) - \sigma' \ln a - \mu' \ln b =$ const. for this deterministic 
dynamics, the solutions to eqs.~(\ref{lvmfeq}) are strictly periodic non-linear 
oscillations that precisely return to the system's initial state. Although popular 
in the fields of ecology and biology, the Lotka--Volterra model is also often 
criticized for being too simplistic and mathematically unstable. This is due to 
several simplifying and likely unrealistic assumptions: First, the prey always have 
a sufficient amount of food available, whence its depletion is neglected, and the
prey's nourishment source is not explicitly represented in the model. Second, the 
only source of food for the predator species is the prey, and its consumption is a
necessary requirement for the predators' reproduction. Third, there is no specified
limit on the prey intake for the predators. Fourth, the rate of change of either
population is directly proportional to its size. Finally, during the temporal
evolution any environmental influence is assumed fixed in time, and crucial 
concepts such as trait inheritance, mutations, or natural selection play no role. 

In our research, we use Monte Carlo simulations for the stochastic Lotka--Volterra 
model based on the reactions~(\ref{lvreac}) performed on a two-dimensional square 
lattice with $512 \times 512$ sites and periodic boundary conditions to fully 
account for emerging spatial structures and internal reaction noise. We note that 
we have also performed simulations on two-dimensional square lattices with 
$256 \times 256$ and $128 \times 128$ sites; aside from overall noisier data, as 
one would expect, we obtain no noticeable quantitative differences. Given that the 
correlation lengths $\xi$ measured below are much smaller than these system sizes, 
this is not surprising. Due to our limited computational resources, we have not 
attempted runs on even larger systems. In the following, all listed Monte Carlo 
data and extracted quantitative results refer to $512 \times 512$ square lattices. 
Also, for the reaction processes, we only consider the four nearest-neighbor sites, 
and have not extended interactions to larger distances. In our model, we implement 
occupation number limitations or finite local carrying capacities; i.e., the number
of particles on any lattice site is restricted to be either $0$, if the site is 
empty, or $1$, if it is occupied by a predator or a prey individual. We shall 
examine the population densities of each species, given by their total particle 
number divided by the number of lattice sites, and aim to quantify the ensuing 
oscillations and through characteristic observables that include their frequency 
and attenuation, as well as typical population cluster sizes as determined by their
spatial correlation length. 

The simulation algorithm for the death, reproduction, and mutual interactions of 
the prey and predator particles proceeds as follows~\cite{rey1986, agf1999, a1999}:
For each iteration, an occupied site is randomly selected and then one of its four 
adjacent sites is picked at random. If the two selected sites contain a predator and
a prey particle, a random number $x_1 \in [0,1]$ is generated; if $x_1 < \lambda$, 
the prey individual is removed and a newly generated predator takes its place. 
Similarly, if the occupant is a predator, a random number $x_2 \in [0,1]$ is 
generated, and the particle is removed if $x_2 < \mu$. Yet if the initially selected
occupant is a prey particle and the chosen neighbor site empty, a random number 
$x_3 \in [0,1]$ is generated; if $x_3 < \sigma$ a new prey individual is added to 
this site. One Monte Carlo Step (MCS) is considered completed when on average all 
particles have participated in the reactions once.

The variables that can be tuned in our simulations are: the system size $L$, the 
initial predator density $\rho_A(0)$, the initial prey density $\rho_B(0)$, the
predator death rate $\mu$, the prey reproduction rate $\sigma$, the predation rate 
$\lambda$, and the number of Monte Carlo steps. We chose the linear system size 
$L = 512$. Naturally one must avoid starting the simulations from one of the 
absorbing states. For any non-zero initial predator and prey density, the population
numbers and particle distribution at the outset of the simulation runs influence the
system merely for a limited time, and the final (quasi-)stationary state of the 
system is only determined by the three reaction rates~\cite{ct2016}. In our 
simulations, the rates $\mu$ and $\sigma$ are kept constant for simplicity, while 
$\lambda$ is considered to be the only relevant control parameter. The dynamical 
properties are generically determined by the ratio of the reaction rates; the 
subsequent results apply also for different sets of $\mu$ and $\sigma$ with 
appropriately altered predation rate $\lambda$. Since we only have two species, 
predators and prey, $0 \leq \rho_A + \rho_B \leq 1$ due to the site occupation 
resctrictions. For each parameter set we use a suitable number of independent 
simulation runs and use the average of these repeats in the data analysis to reduce
statistical errors.

\begin{figure*}[t]
\centering
\includegraphics[width=0.5\columnwidth,height=0.5\columnwidth]{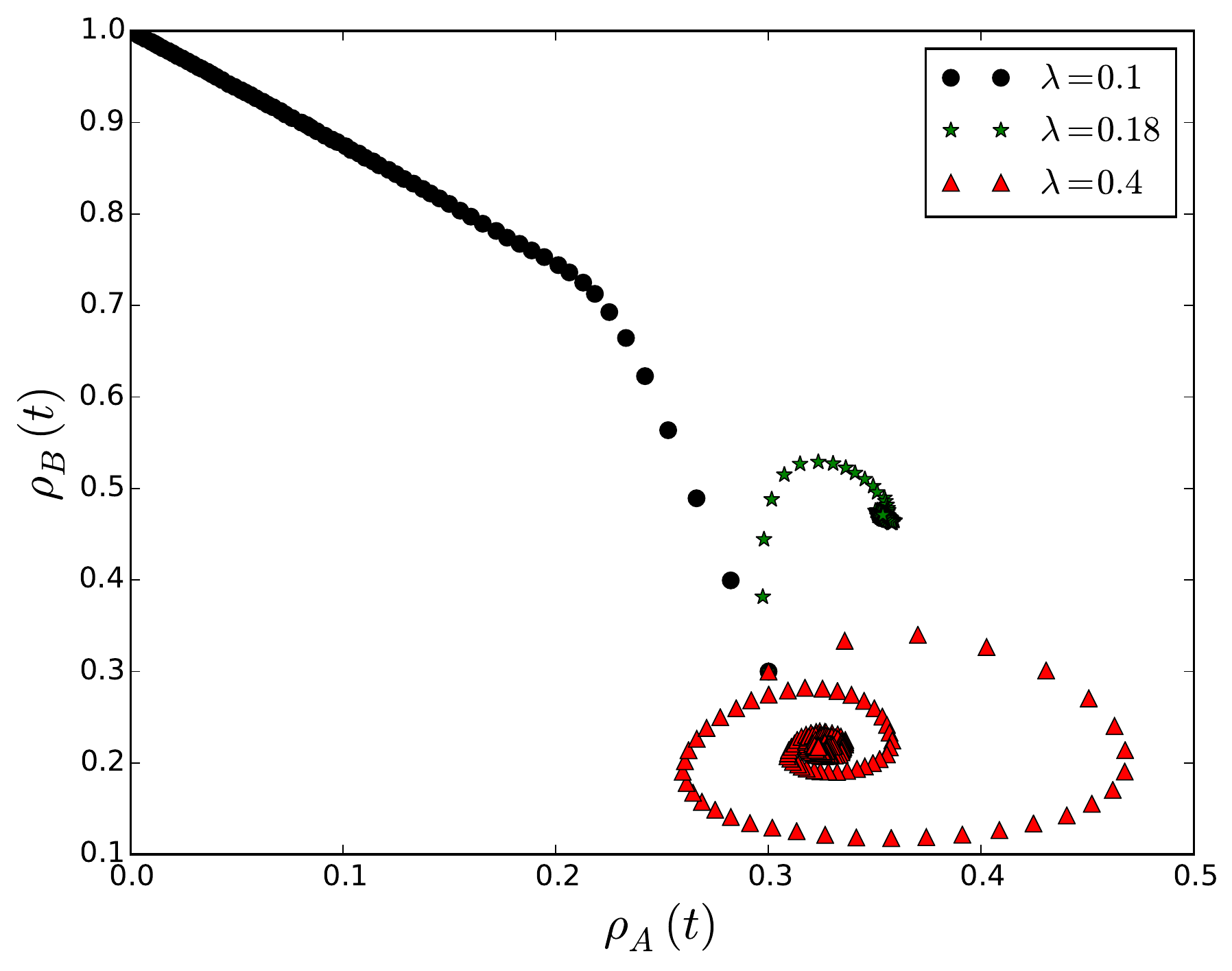}  
\caption{Monte Carlo simulation trajectories for a stochastic Lotka--Volterra model 
        on a $512 \times 512$ square lattice with periodic boundary conditions and 
        restricted site occupancy shown in the predator $\rho_A(t)$ versus prey 
        density $\rho_B(t)$ phase plane ($\rho_A(t) + \rho_B(t) \leq 1$) with 
        initial values $\rho_A(0) = 0.3 = \rho_B(0)$, fixed reaction probabilities 
        $\mu = 0.125$, $\sigma = 1.0$, and different predation efficiencies:  
        (i) $\lambda = 0.1$ (black dots): predator extinction phase; 
        (ii) $\lambda=0.18$ (green stars): direct exponential relaxation to the 
        quasi-stationary state just above the extinction threshold in the 
        predator-prey coexistence phase; 
        (iii) $\lambda = 0.4$ (red triangles): the trajectory spirals into a stable
        fixed point, signifying damped oscillations deep in the coexistence phase.}
\label{fig1}
\end{figure*} 
In the case of very low predation rates $\lambda$, the predators will gradually 
starve to death, and the remaining prey will finally occupy the whole system. On 
the other hand, when $\lambda$ is large, there is a finite probability (in any 
finite lattice) that all prey individuals would be devoured; subsequently the 
predators would die out as well because of starvation. In fact, the absorbing 
extinction state is the only truly stable state in a finite population with the 
stochastic dynamics (\ref{lvreac}). However, in sufficiently large systems, 
quasi-stable states in which both species survive with relatively constant 
population densities during the entire simulation duration are indeed observed in 
certain regions of parameter space. Fig.~\ref{fig1} shows three single-run 
simulation results, plotting the prey population density $\rho_B(t)$ versus that 
of the predators $\rho_A(t)$ with the reaction probabilities $\mu = 0.125$ and 
$\sigma = 1.0$ held fixed; we thus select the non-linear predation reaction rate 
$\lambda$ as the only control parameter. We chose the initial population densities
as $\rho_A(0) = 0.3 = \rho_B(0)$ with the particles randomly distributed among the 
lattice sites. With $\lambda = 0.1$ (black dots), the predators have low predation 
efficiency and thus gradually go extinct; the system then reaches an absorbing 
state with only prey particles remaining and ultimately filling the entire lattice 
($\rho_B \to 1$). If we increase the value of $\lambda$ to $0.18$ (green stars), 
just above the predator extinction threshold, the system relaxes exponentially to 
a quasi-stationary state with non-zero densities for both species. For 
$\lambda = 0.4$ (red triangles), the system resides deep in this coexistence phase 
and the simulation trajectory spirals into a stable fixed point, indicating damped 
oscillatory kinetics. According to our investigations, we estimate the critical 
predation rate of the predator extinction phase transition point at 
$\lambda_c = 0.12 \pm 0.01$.

\section{Boundary Effects at a Coexistence / Predator Extinction Interface}

\begin{figure*}[t]
\centering
\includegraphics[width=\columnwidth,height=0.42\columnwidth]{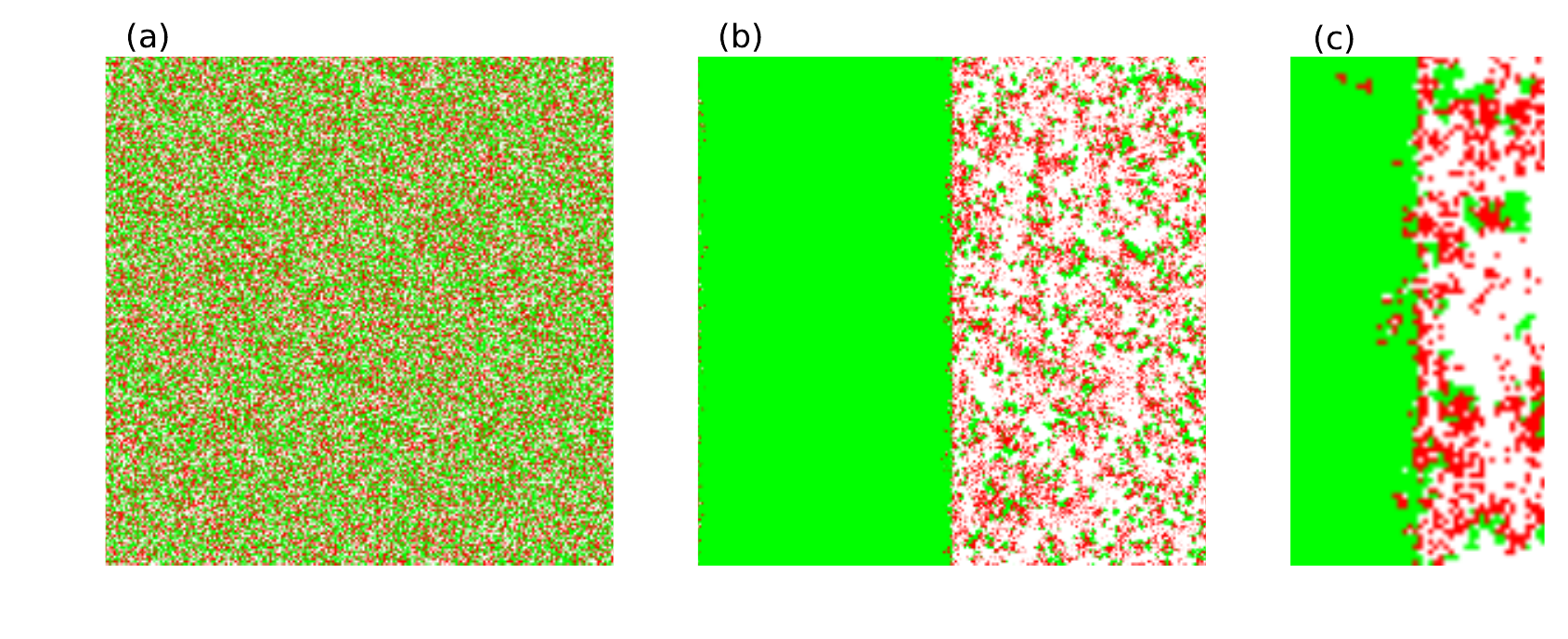}  
\caption{Snapshots of the spatial particle distribution on a $512 \times 512$ 
	    lattice (with periodic boundary conditions) that is split into equally
        large predator extinction (left) and species coexistence (right) regions:
        prey are indicated in green, predators in red, white spaces in white.
        (a) Random initial distribution with densities $\rho_A = 0.3 = \rho_B$; 
        (b) state of the system after $1000$ MCS, when it has reached a 
        quasi-stationary state with uniform rates $\mu = 0.125$ and $\sigma = 1.0$,
        while $\lambda_l = 0.1$ on columns $[0, 255]$, $\lambda_r = 0.8$ on columns
        $[256, 511]$; (c) close-up of a local $100 \times 50$ area at the boundary.}
\label{fig2}
\end{figure*}
Natural environments vary in space and boundaries are formed between different 
regions, yielding often quite sharp interfaces, e.g., between river and land, 
desert and forest, etc. At the boundaries of such spatially inhomogeneous systems,
interesting phenomena may arise. In order to study boundary effects on simple
predator-prey population dynamics, we split our simulation domain into two equally
large pieces with one half residing in the predator extinction state, and the 
other half in the two-species coexistence phase. We use a two-dimensional lattice 
with $512 \times 512$ sites with periodic boundary conditions, and index the 
columns with integers in the interval $[0, 511]$. Whereas the predator death and 
prey reproducation rates are uniformly set as $\mu = 0.125$ and $\sigma = 1.0$ on 
all sites, we assign $\lambda_l = 0.1 < \lambda_c$ on columns $[0, 255]$ to 
enforce predator extinction on the ``left'' side, and $\lambda_r = 0.8 > \lambda_c$
for the columns on the ``right'' half with indices $[256, 511]$, which is thus 
held in the predator-prey coexistence state. Fig.~\ref{fig2}(a) depicts the 
initial random particle distribution with equal population densities 
$\rho_A = 0.3 = \rho_B$. After the system has evolved for $1000$ MCS, a 
quasi-steady state is obtained as shown in Fig.~\ref{fig2}(b), and the close-up 
near the boundary (c). The predators are able to penetrate into the ``left'' 
absorbing region by less than $10$ columns, and no predator individuals are 
encountered far away from the active-absorbing interface. On the right half, we 
observe a predator-prey coexistence state with the prey particles forming clusters 
surrounded by predators and predation reactions occurring at their perimeters.

\begin{figure*}[t]
\centering
\includegraphics[width=0.7\columnwidth,height=0.45\columnwidth]{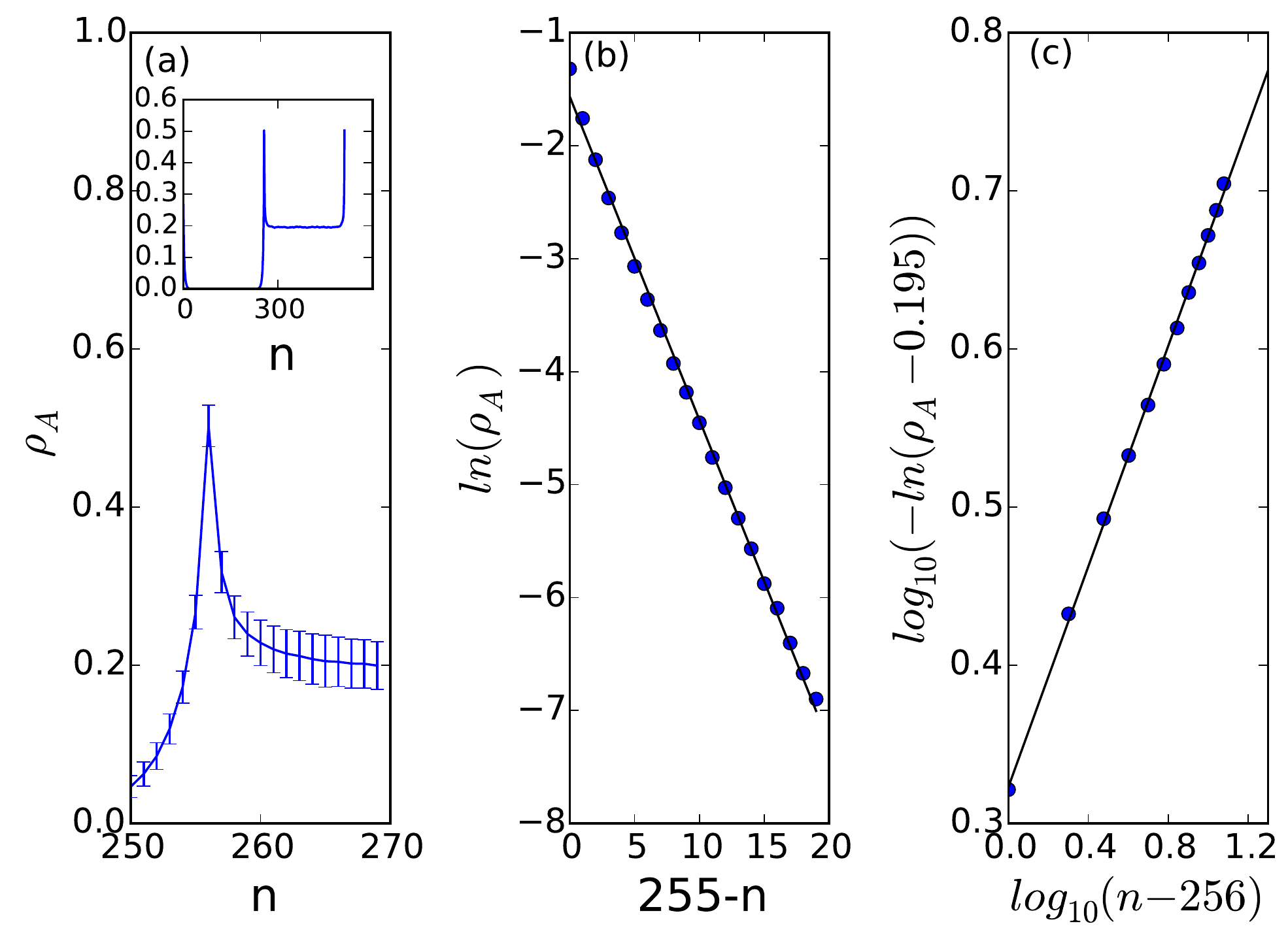} 
\caption{After the split system with rates $\mu = 0.125$, $\sigma = 1.0$ and 
        $\lambda_l = 0.1$ on columns $[0, 255]$, $\lambda_r = 0.8$ on columns 
        $[256, 511]$ evolves for $1000$ MCS, it arrives at a quasi-stationary 
	    state: (a) the main plot shows the column densities of the predator 
        population $\rho_A(n)$ as function of column index $n \in [251, 270]$ with 
		the error bars indicating the standard deviation, and the inset on all 
		$L = 512$ columns (data averaged over $1000$ independent runs); 
	    (b) exponential decay of $\rho_A(n)$ from the boundary (located at 
        $n = 255$) into the absorbing region with $n \in [235, 255]$: The blue 
        dots depict our simulation results, while the black straight line 
        represents a linear regression of the data with slope $k = -0.286$; 
        (c) the column density $\rho_A$ decays to a positive constant value 
        $0.195 \pm 0.001$ deep in the right coexistence region. The blue dots 
        display $\log_{10} (-\ln (\rho_A-0.195))$ versus $\log_{10}(n - 256)$, 
	while the black straight line with slope $l = 0.348$ is obtained from 
	linear data regression.}
\label{fig3}
\end{figure*}

Since only the predator species is subject to the extinction transition into an 
absorbing state, while the prey can survive throughout the entire simulation 
domain, we concentrate on boundary effects affecting the predator population. We 
measure the column densities of predators $\rho_A(n)$, defined as the number of 
predators on column $n$ divided by $L = 512$, and record their averages from 
$1000$ independent simulation runs as a function of column index $n$. As shown in
the inset of Fig.~\ref{fig3}(a), $\rho_A(n)$ decreases to $0$ deep inside the 
absorbing half of the system, and reaches a positive constant $0.195 \pm 0.001$ 
within the active region. The main graph focuses on the boundary region, where we 
observe a marked predator density peak right at the interface (column $256$). The 
predator density enhancement at the boundary is obviously due to the net intrusion 
flow of species $A$ from the active subdomain with high predation rate into the 
predator extinction region with abundant food in the form of the near uniformly 
spread prey population. We also ran simulations for other predation rate pairs 
such as $\lambda_l = 0.1$ and $\lambda_r = 0.2$ (still in the coexistence phase), 
and observed very similar behavior (except that the peak of $\rho_A$ appeared on 
column $257$ in that situation instead of at $n = 256$). 

Fig.~\ref{fig3}(b) shows the exponential decay of the predator column density 
$\rho_A(n)$ as function of the distance $|255 - n|$ from the boundary (located at 
$n = 255$) towards the ``left'', absorbing side. A simple linear regression gives 
the inverse characteristic decay length $k = -0.286$. However, on the ``right''
active half of the system, $\rho_A(n)$ neither fits exponential nor algebraic 
decay. Instead, $\rho_A$ reaches the asymptotic constant value $0.195 \pm 0.001$ 
deep in the coexistence region through an apparent stretched exponential form
$\rho_A(n) \sim e^{-(n-256)^l} + 0.195$ with stretching exponent $l \approx 0.348$,
as demonstrated in Fig.~\ref{fig3}(c).  

On the ``right'' semidomain set in the predator-prey coexistence phase, the 
particle reproduction processes induce clustering of individuals from each species.
The cluster size may vary with the distance from the boundary. We utilize the
correlation length $\xi$, obtained from the equal-time correlation function $C(x)$,
to characterize the spatial extent of these clusters. For species 
$\alpha, \beta = A, B$, the (connected) correlation functions are defined as 
$C_{\alpha \beta}(x) = \langle n_\alpha(x) \, n_\beta(0) \rangle - \langle 
n_\alpha(x) \rangle \, \langle n_\beta(0) \rangle$, where $n_\alpha(x) = 0, 1$ 
denotes the local occupation number of species $\alpha$ at site $x$ \cite{mmu2007}.
For $x = 0$ and $\alpha = \beta$, in a spatially homogeneous system it is simply 
given by the density $\langle n_A \rangle$: 
$C_{\alpha \alpha}(0) = \langle n_A \rangle (1 - \langle n_A \rangle)$. For 
$|x| > 0$, $\langle n_\alpha(x) \, n_\beta(0) \rangle$ is computed as follows: 
First choose a site, and then a second site at distance $x$ away from the first 
one. $n_\alpha(x) \, n_\beta(0)$ equals $1$ only if the first site is occupied by 
an individual of species $\beta$, and the second one by an particle of species 
$\alpha$, otherwise the result is $0$. One then averages over all sites.

\begin{figure*}[t]
\centering
\includegraphics[width=0.7\columnwidth,height=0.45\columnwidth]{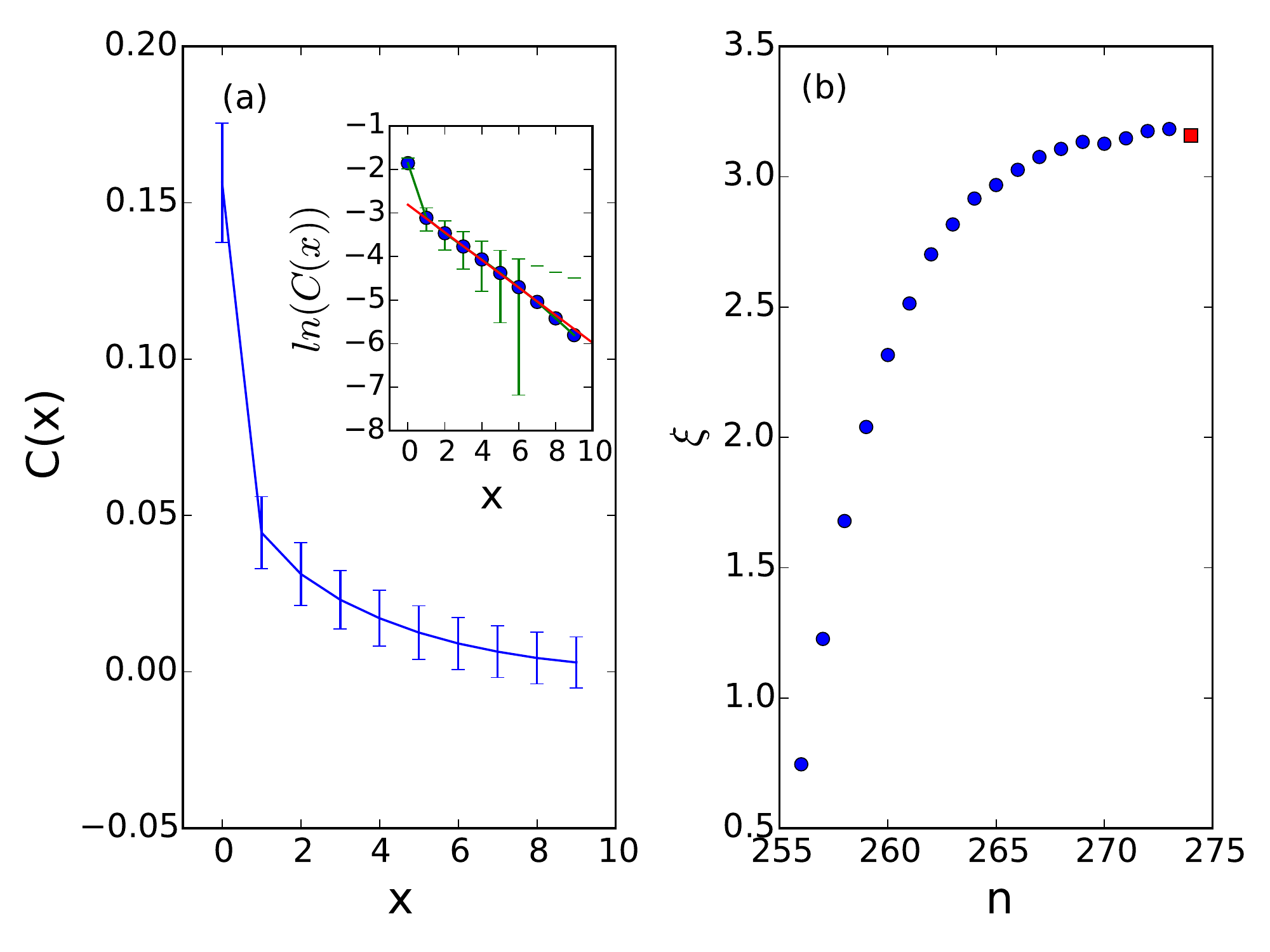} 
\caption{(a) Main panel: the predator correlation function $C_{AA}(x,n)$ on column 
        $n = 274$ (data averaged over $1000$ independent simulation runs). Inset:
        $\ln(C_{AA}(x,n))$; the red straight line indicates a simple linear 
        regression of the data points with $x \in [1,6]$, and yields the 
        characteristic decay length $\xi(n = 274) \approx 3.2$; the error bars 
		indicate the standard deviation.
	    (b) Correlation length $\xi(n)$ versus column number $n$, with $\xi(n)$ 
        defined as the negative reciprocal of the slope of $\ln(C_{AA}(x,n))$.}
\label{fig4}
\end{figure*}
Here, we compute the predator correlations $C_{AA}(x,n)$ on a given column $n$, 
i.e., we only take the mean in the above procedure over the $L = 512$ sites on 
that column. The main panel in Fig.~\ref{fig4}(a) shows the predator correlation 
function $C_{AA}(x)$ on column $n = 274$ with $x \in [0, 9]$, where the $C_{AA}(x)$
gradually decreases to zero. The inset presents the same data in a logarithmic 
scale, demonstrating exponential decay according to $C(x,n) \sim e^{- x / \xi(n)}$. 
Since the statistical errors grow at large distances $x$, we only use the initial 
data points up to $x = 6$ for the analysis. Linear regression of 
$\ln(C_{AA}(x,n = 274))$ over $x \in [1, 6]$ gives $\xi(n = 274) \approx 3.2$, 
indicated as red square in Fig.~\ref{fig4}(b). In the same manner, we obtain the 
characteristic correlation lengths $\xi(n)$ for each column $n$ as shown in 
Fig.~\ref{fig4}(b), starting at the interface at $n = 256$. We observe $\xi(n)$ to 
increase by about a factor of four within the first ten columns away from the 
boundary, and then saturate at the bulk value $\xi \approx 3.2$. Near the absorbing
region, the predator clusters are thus much smaller, owing to the net flux of 
predators across the boundary into the extinction domain. These values of $\xi$ are 
measured after the entire system has reached its (quasi-)steady state after $1000$ 
MCS, and would not change for longer simulations run times. We note that the 
relationship between the correlation length $\xi$ and the predation rate $\lambda$ 
is manifestly not linear, i.e., a very large value of $\lambda$ does not imply huge
predator clusters. We surmise that the cluster size remains finite even in that 
scenario, and the predators would penetrate into the ``left'' absorbing region for 
a finite number of columns only. For sufficiently large domain size, the system 
should thus remain spatially inhomogeneous even for very high predation rates 
$\lambda$. Finally, the dependence of the typical cluster size $\xi(n)$ on column 
index $n$ correlates inversely with the column density plotted in Fig.~\ref{fig3}: 
High local density corresponds to small cluster size and vice versa. We note that 
the product $\rho_A(n) \, \xi(n)$ is however not simply constant across different 
columns; rather it is minimal near the boundary (at $n = 256$), then increases away
from the interface, and ultimately reaches a fixed value within $10$ columns inside 
the active region.

\begin{figure*}[t]
\centering
\includegraphics[width=0.5\columnwidth,height=0.4\columnwidth]{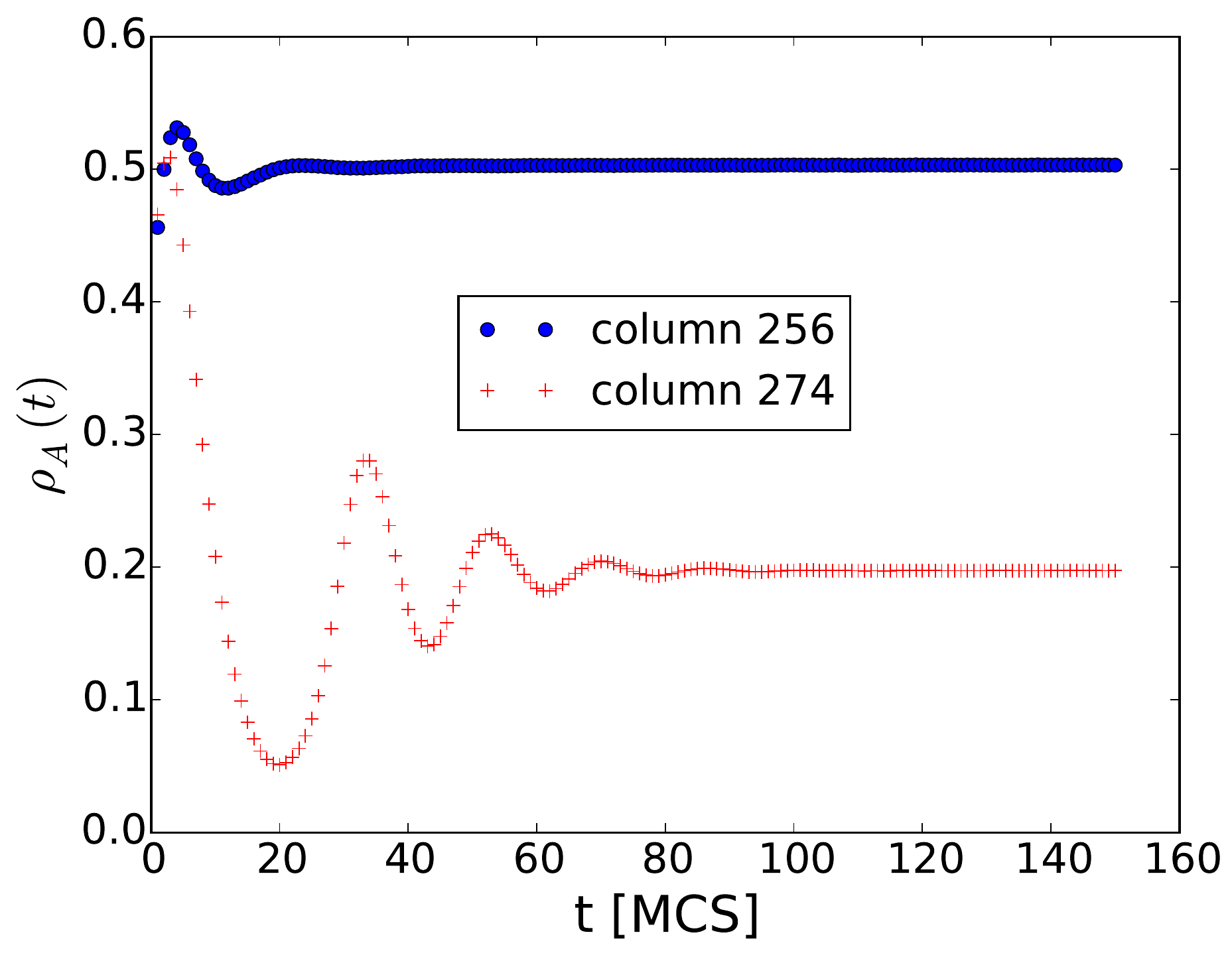}  
\caption{The temporal evolution of the average predator column densities 
        $\rho_A(t,n)$ (averaged over $1000$ independent runs) on columns $n = 256$ 
        (blue dots) and $n = 274$ (red plus marks), with initial predator density 
	    $\rho_A(0) = 0.3$ and rates $\mu = 0.125$, $\sigma = 1.0$, 
        $\lambda_l = 0.1$ on columns $n \in [0, 255]$, and $\lambda_r = 0.8$ for 
        $n \in [256, 511]$.}
\label{fig5}
\end{figure*} 
\begin{figure*}[t]
\centering
\includegraphics[width=0.6\columnwidth,height=0.45\columnwidth]{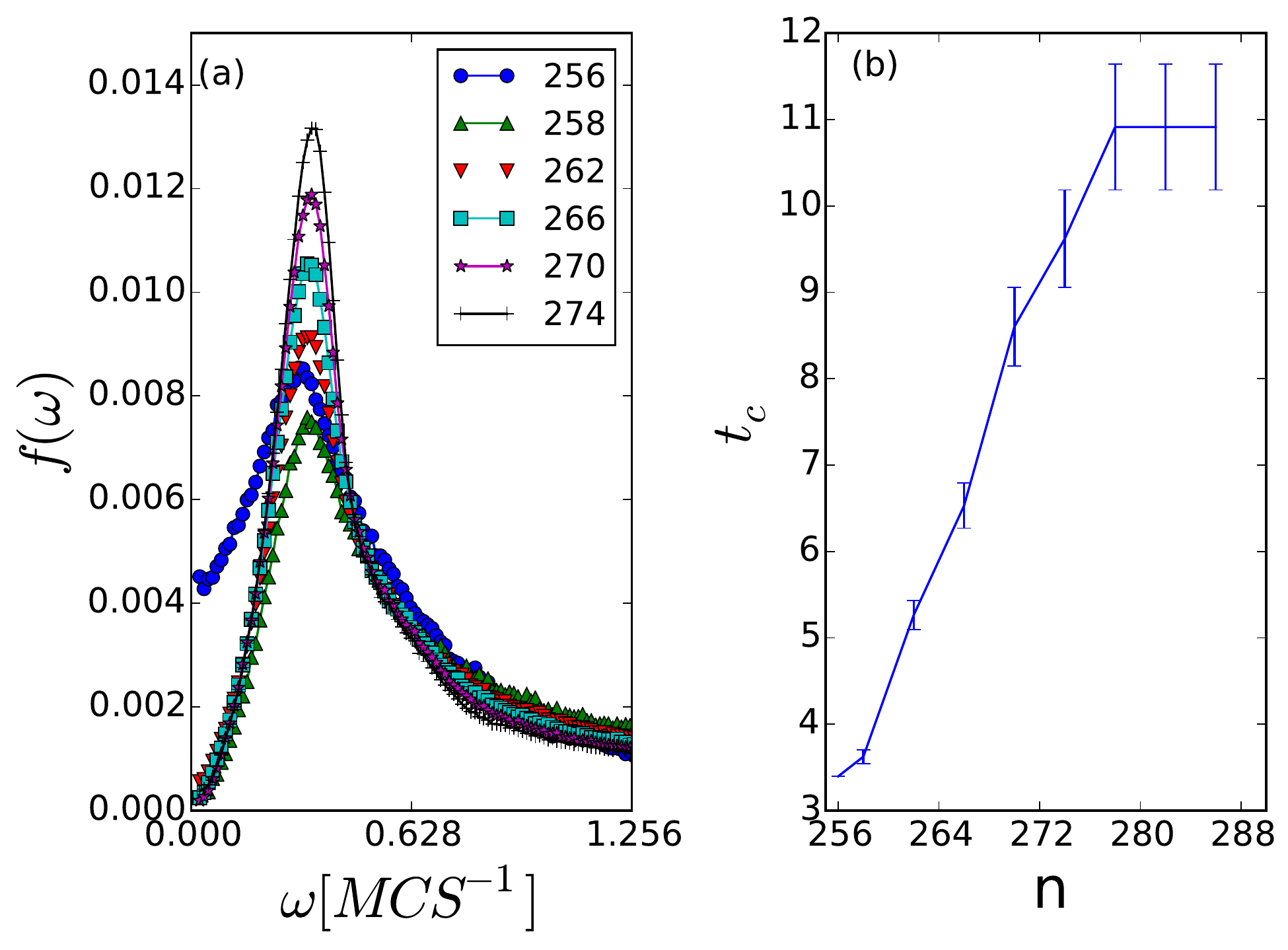}  
\caption{(a) Fourier transform amplitude $f_A(\omega,n)$ of the predator column 
        density time evolution on columns $n = 256$ (blue dots), $258$ (green 
		triangles up), $262$ (red triangles down), $266$ (cyan squares), $270$ 
		(magenta stars), and $274$ (black plus marks), with rates $\mu = 0.125$, 
		$\sigma = 1.0$, and $\lambda_l = 0.1$ for $n \in [0, 255]$, 
		$\lambda_r = 0.8$ for $n \in [256, 511]$; 
        (b) measured characteristic decay time $t_c(n)$ on columns near the 
        active-absorbing boundary, inferred from the peak widths in (a), with the 
		error bars representing the standard deviation.}
\label{fig6}
\end{figure*}
Spatially homogeneous stochastic Lotka--Volterra systems display damped population
oscillations in the predator-prey coexistence phase after being initialized with
random species distribution, see, for example, the (red triangle) trajectory in 
Fig.~\ref{fig1} for predation rate $\lambda = 0.4$. We next explore the boundary 
effects on these population oscillations near the active-absorbing interface. We 
prepare the system with the same parameters as mentioned above so that its 
``left'' half is in the absorbing state while the ``right'' side sustains species 
coexistence. The initial population densities are again set to 
$\rho_A(0) = 0.3 = \rho_B(0)$, with the particles randomly distributed on the 
lattice. We then measure the column predator densities as a function of time (MCS).
Fig.~\ref{fig5} displays the temporal evolution of $\rho_A(t,n)$ on columns 
$n = 256$ and $n = 274$. We observe the oscillations on the column closest to the 
interface to be strongly damped, whereas deeper inside the active region the 
population oscillations are more persistent and subject to much weaker attenuation.
Both column densities asymptotically reach the expected quasi-steady state values. 

In order to determine the dependence of the local oscillation frequencies on the 
distance from the active-absorbing interface, we compute the Fourier transform 
amplitude $f_A(\omega,n) = | \int e^{- i \omega t} \, \rho_A(t,n) \, dt |$ of the 
column density time series data  by means of the fast Fourier transform algorithm 
for $n \in [256, 274]$, as shown in Fig.~\ref{fig6}(a). Assuming the approximate
functional form $\rho_A(t,n) \sim e^{- t / t_c(n)} \, \cos(2 \pi t / T(n))$, we 
may then identify the peak position of $f_A(\omega,n)$ with the characteristic 
oscillation frequency $2 \pi / T(n)$, and the peak half-width at half maximum with 
the attenuation rate or inverse relaxation time $1 / t_c(n)$. We find that the 
oscillation frequencies are constant except for the column at the boundary
($n = 256$), which shows a very slight enhancement. We conclude that the presence 
of the extinction region does not markedly affect the frequency of the population 
oscillations in the active regime. In contrast, the attenuation rate increases by a
factor of three within about $20$ columns in the vicinity of the interface, as 
demonstrated in Fig.~\ref{fig6}(b). Beyond $n \approx 278$ in the coexistence 
region, the relaxation time assumes its constant bulk value.

\section{Checkerboard Division of the System}

\begin{figure*}[t]
\centering
\includegraphics[width=1.0\columnwidth,height=0.35\columnwidth]{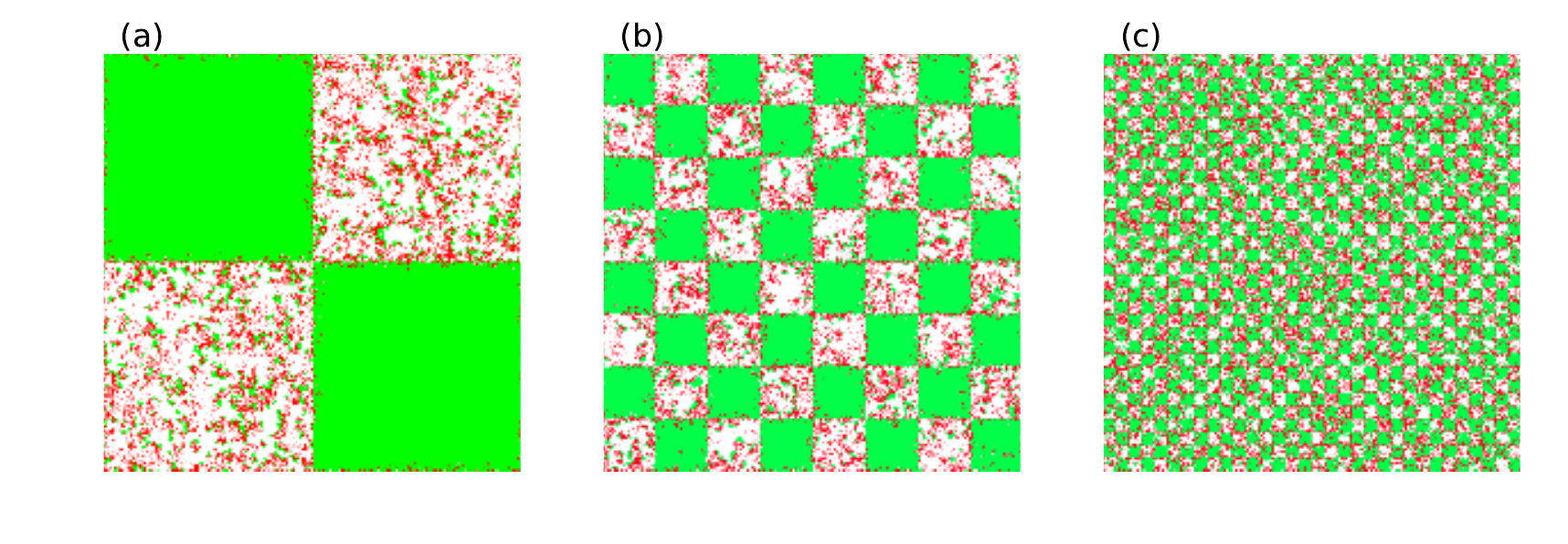}  
\caption{Snapshots of the distribution of predator (red) and prey (green) particles 
        after the system has evolved for 1000 MCS with rates $\mu = 0.125$, 
        $\sigma = 1.0$, and $\lambda$ switched alternatingly between the values 
        $0.1$ (predator extinction) and $0.8$ (species coexistence) on neighboring 
        subdomains, as the full $512 \times 512$ system is periodically divided 
        into successively smaller square patches with lengths $256$ (a), $64$ (b), 
        and $16$ (c), respectively. The square subdomains dominantly colored in 
        green reside in the extinction state ($\lambda = 0.1$), whereas 
        predator-prey coexistence pertains to the other patches ($\lambda = 0.8)$.}
\label{fig7}
\end{figure*} 
To further explore boundary (and finite-size) effects in spatially inhomogeneous
Lotka--Volterra systems, we proceed by successively dividing the simulation domain
into subdomains in a checkerboard pattern, setting the predation rate to two 
distinct values in neighboring patches, and thus preparing them alternatingly in 
either the active coexistence or absorbing predator extinction states. 
Fig.~\ref{fig7}(a) shows a case when the system is split into four subregions with 
$\sigma = 1.0$ and $\mu = 0.125$, and with two distinct values for the predation 
rate $\lambda = 0.1$ and $0.8$ assigned to alternating patches of the $2 \times 2$
checkerboard structure. Note that the low predation rate value posits the 
corresponding patches in the predator extinction state, whereas the subdomains 
with the high predation rate reside in the species coexistence phase. 
Figures~\ref{fig7}(b) and (c) depict the situations when the total simulation 
domain with $512 \times 512$ sites is respectively split into $8 \times 8$ and 
$32 \times 32$ square patches: If for a given box $\lambda$ is set to $0.1$, then 
the adjacent square subdomains above, below, to its right, and to its left are 
given a value $\lambda = 0.8$.

$1000$ simulations were performed for each setting, and the averages over these 
independent runs were used to analyze the data. We also generated and inspected 
simulation videos: snapshots are depicted in Fig.~\ref{fig7}. As we split the 
system into successively smaller and more pieces in the checkerboard-patterned 
fashion with $\lambda$ switching between $0.1$ and $0.8$ on neighboring subregions,
we find the boundaries to have less of an impact on the population densities. We 
observe that in this sequence the prey density decreases on the patches with lower 
predation rate $0.1$, but stays roughly the same on the subdomains where 
$\lambda = 0.8$. The predator density in contrast increases in both the active 
and absorbing regions as the subdivision proceeds. We have also confirmed that 
these changes in the total population densities naturally become less significant 
if the two different predation rate values are chosen closer to each other.

\begin{figure*}[t]
\centering
\includegraphics[width=0.45\columnwidth,height=0.45\columnwidth]{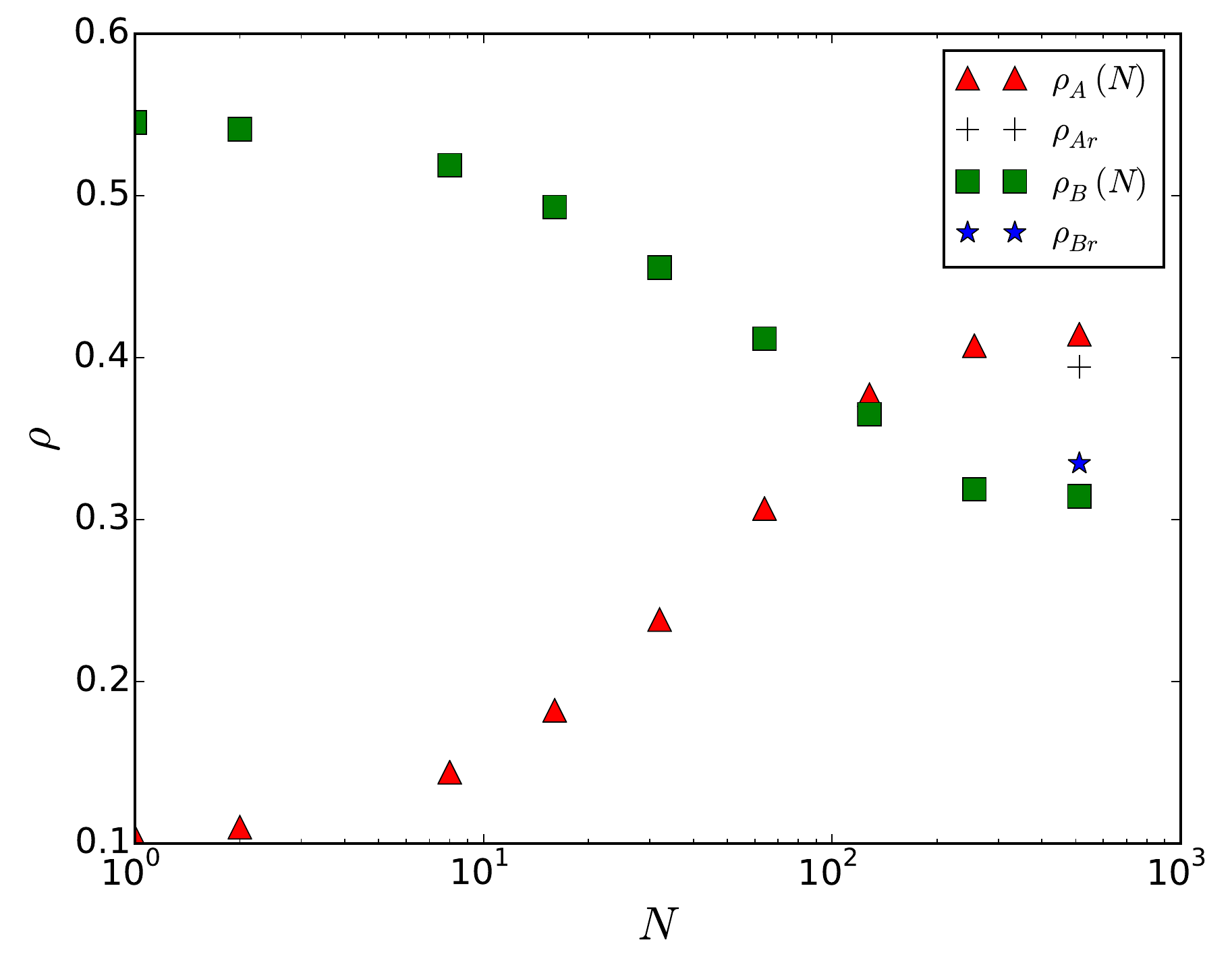}  
\caption{Total population densities for predators (red triangles) and prey (green 
	squares) versus number of checkerboard-patterned subdivisions $N$ of the 
	simulation domain, after the system has evolved for $1000$ MCS and reached 
        a quasi-stationary state, with reaction rates $\mu = 0.125$, 
        $\sigma = 1.0$, and $\lambda$ alternatingly switched between $0.1$ and 
        $0.8$. For comparison, the graph also shows the total quasi-steady state 
        population densities for predators (black plus) and prey (blue star) in a 
        system with randomly assigned predation, drawn with equal probability from 
        a bimodal distribution with values $\lambda = 0.1$ and $0.8$.}
\label{fig8}
\end{figure*} 
In Fig.~\ref{fig8}, we plot the total (summed over all subdomains) predator and 
prey population densities $\rho$ in the simulation domain split into $N \times N$
checkerboard patches, as functions of $\log_{10} N$. Here, $N = 1$ corresponds to 
the situation studied in section~3, where the system was divided into two 
rectangular subdomains. The other values of $N = 512, 256, 128, 64, 32, 16, 8, 2$ 
refer to checkerboard square patches with lengths $512 / N$. The mean population 
density $\rho$ shown for each data point represents an average of $1000$ 
independent simulation runs; the associated statistical error was very small, with
a standard deviation of order $10^{-3}$. As apparent in the data, the overall 
population predator density $\rho_A$ monotonically increases with growing number 
$N$ of subdivisions, while the prey density $\rho_B$ decreases.

We also performed the analogous sequence of measurements for other pairs of
predation rate values. For instance, with checkerboard subdomains with 
$\lambda = 0.1$ and $0.2$ (also just within the species coexistence range, see 
Fig.~\ref{fig1}), the population density changes with increasing $N$ are less
pronounced than in Fig.~\ref{fig8}, and $\rho_A$, $\rho_B$ acquire maximum and
minimum values at $N = 256$ rather than $512$. The origin of this slight shift can
be traced to the fact that the predator correlation length is of order one lattice
constant at the boundary of the $\lambda = 0.1 / 0.8$ system, but extends over 
about two sites for the $0.1 / 0.2$ case. 
 
For comparison, we also measured the overall population predator and prey 
densities in a Lotka--Volterra system with quenched spatial disorder in the 
predation rates, where either of the two values $\lambda = 0.1$ and $0.8$ are
assigned at random to each lattice site with equal probability. The resulting net
population density values are also shown in Fig.~\ref{fig8}; they are close, but
not identical to those obtained for the $N = 512$ system, for which these two
predation rates are alternatingly assigned to the lattice sites in a periodic
regular manner. We would expect the population densities in these two distinct
systems to reach equal values if the associated correlation lengths at the
boundaries were large compared to the lattice constant, which is however not the
case here.

\section{Conclusion}

In this work, we have focused on studying boundary effects in a stochastic 
Lotka--Volterra predator-prey competition model on a two-dimensional lattice, by
means of detailed Monte Carlo simulations. We first considered a system split into
two equally large parts with distinct non-linear predation rates, such that one 
domain is set to be in the predator extinction state, while the other one resides 
in the two-species coexistence phase. We have primarily addressed the influence of
such an absorbing-active separation on both populations' density oscillations as 
function of the distance from the boundary. 

We find a remarkable peak in the column density oscillation amplitude of the 
predator population, as shown in Fig.~\ref{fig3}(a), which reflects its net steady 
influx towards the absorbing region. Correspondingly, the predator correlation 
length that characterizes the typical cluster size reaches a minimum value at the 
boundary, see Fig.~\ref{fig4}(b). The population oscillation frequency there shows 
only small deviations from its bulk value, while the attenuation rate is locally 
strongly enhanced, see Fig.~\ref{fig6}(b), inducing overdamped relaxation 
kinetics. Overall, the ecosystem remains stable. 

Furthermore, upon splitting the system successively into more pieces in a 
checkerboard fashion, the observed boundary effects become less significant, and 
as demonstrated in Fig.~\ref{fig8}, the overall population densities acquire 
values that are close to those in a disordered system with randomly assigned 
predation rates drawn from a bimodal distribution.

\end{document}